\documentstyle[aps,prl,epsf,rotate]{revtex}
\begin{document}

\twocolumn[\hsize\textwidth\columnwidth\hsize\csname
@twocolumnfalse\endcsname

\title{Field-tilt Anisotropy Energy in Quantum Hall Stripe States}


\author{
T. Jungwirth$^{1,2}$, 
A. H. MacDonald$^{1}$, 
L. Smr\v{c}ka$^{2}$,
and
S. M. Girvin$^{1}$} 
\address{$^{1}$Department of Physics,
Indiana University, Bloomington, Indiana 47405}
\address{$^{2}$Institute of Physics ASCR,
Cukrovarnick\'a 10, 162 00 Praha 6, Czech Republic}
\date{\today}
\maketitle

\begin{abstract}

Recently reported giant anisotropy in the longitudinal resistivity of 
a 2D electron system with valence Landau level index $N \ge 2$ 
has been interpreted as a signal of unidirectional
charge density wave (UCDW) ground states.  
We report on detailed Hartree-Fock calculations of the UCDW orientation 
energy induced by a tilted magnetic field.
We find that for current experimental samples stripes are oriented
perpendicular to the in-plane field, consistent with experiment.
For wider 2D electron systems we predict tilt-induced 
stripe states with variable anisotropy energy sign.

\end{abstract}



\vskip2pc] 

Several groups\cite{lilly,lilly2,duetal,panetal,shayegan} 
have reported the observation of strong anisotropies and
nonlinearities in the low temperature magnetotransport of clean
2D electron systems over ranges of Landau level (LL) filling factor 
surrounding $\nu = n +1/2$ for $n \ge 4$, {\it i.e.}, for 
valence LL orbital index $N \ge 2$. 
Although the origin of these anomalies
has not been firmly established, the anisotropy is 
probably associated with the UCDW states
which have been predicted to occur under precisely these circumstances
\cite{koukalov,moessner}.
Recently Pan {\em et al.}\cite{panetal} and Lilly {\em et al.}\cite{lilly2} 
have discovered that the isotropic gapped $\nu=5/2$ quantum Hall state 
gives way to the anisotropic state for sufficiently large in-plane magnetic
fields.  
Shayegan and Manoharan\cite{shayegan} have observed that in a 
2D hole system the $\nu=5/2$ state is already
anisotropic even without in-plane field, indicating that lower electron density
(more LL mixing) can stabilize the CDW.  Both of these observations 
are consistent with an anisotropic spontaneously-broken
orientational-symmetry state, like the UCDW state.  
Several recent theoretical papers\cite{fradkin,fertig,rezayi}
have explored the properties of these `liquid crystal'  
states for perpendicular field. 

In this paper we evaluate the dependence of 
UCDW state's energy on its orientation relative 
to the in-plane field component, when the magnetic field
is tilted away from the 2D electron system normal.   
Theoretical studies along similar lines have recently
been carried out by two other groups\cite{phillips,fogler}.
We find that screening due to polarization of remote LL's plays
an essential role for the preferred orientation of the stripes.
Using a realistic model for the sample of Lilly {\em et al.}
(a single GaAs/Al$_x$Ga$_{1-x}$As heterojunction with density $N_e=2.67\times
10^{11}$ cm$^{-2}$) we quantitatively determine the
anisotropy energy and find that the stripes prefer to be aligned 
perpendicular to the in-plane field for the whole range of studied
field-tilt angles and filling factors, consistent with
experimental observations.  The same conclusion was found to apply for the
lower density ($N_e=2.2\times10^{11}$ cm$^{-2}$) sample  of Pan {\em et al.}.
To explore the dependence of this result on system geometry we have 
repeated these calculations for a parabolic confinement quantum well
models with variable subband separation.  These calculations 
reveal a mechanism for tilt-induced UCDW states in samples with 
more than one occupied subband for which the perpendicular field
state is expected to be isotropic.  We find that 
stripe orientation parallel to the in-plane field is possible
when two subbands are occupied at zero tilt angle.

Our calculation starts from the following observation\cite{other,caveats}.
The property that states within a LL are related to
each other by operations of the magnetic translation group implies
equivalence of any LL to the lowest LL
of a zero-thickness 2D electron
system with a suitably adjusted effective electron-electron interaction.
For the example of interest here, a quasi-2D electron system
in the $x-y$ plane with the magnetic field tilted away from the normal to
the plane\cite{nickila}, we choose the in-plane component $B_{\parallel}$
of the magnetic field to be in the 
$\hat{x}$-direction and use the following Landau gauge for the vector
potential, $\vec{A}=(0,B_{\perp}x-B_{\parallel}z,0)$. 
The one-particle orbitals for any $z$-dependent single-particle
confining potential  can then be written as 
\begin{equation}
\label{wf}
<\vec{r}|k,i,\sigma>=\frac{e^{iky}}{\sqrt{L_y}}\, 
\varphi_{i,\sigma}(x-\ell^2k,z)\; ,
\end{equation}
where $k$ is the wave vector which labels states within LL $i$, 
$\sigma$ is the spin index, and 
$\ell^2 = \hbar c /e B_{\perp}$.  The translational symmetry responsible
for LL degeneracy leads to a 2D wavefunction 
$\varphi_{i,\sigma}(x,z)$
which is independent of the state label $k$, except for the rigid shift
by $\ell^2k$ along $x$-axis.
This in turn leads to two-particle matrix elements of the
Coulomb interactions with a dependence on state labels which
is identical to that for the lowest
LL of a zero-thickness 2D electron system provided the 
2D Coulomb interaction is replaced by
the following effective interaction:  
\begin{equation}
\label{vef}
V(\vec{q})=\frac{4 \pi e^2}{\epsilon}
e^{q^2\ell^2/2}\, \int_{-\infty}^{\infty}
\frac{dq_z}{2\pi}\frac{\left|M_{\sigma}^{i,i}(\vec{q})\right|^2}{q^2+q_z^2}
\end{equation}
where $\vec{q}=(q_x,q_y)$,
$\epsilon_0$ is the semiconductor dielectric function and  
\begin{eqnarray}
\label{mq}
M_{\sigma}^{i,i}(\vec{q})&=&\int_{-\infty}^
{\infty}dx\int_{-\infty}^{\infty}dz 
e^{iq_xx}e^{iq_zz}\nonumber\\
&\varphi_{i,\sigma}&(x+\ell^2q_y/2,z)
\varphi_{i,\sigma}(x-\ell^2q_y/2,z)\; .
\end{eqnarray}

Since the stripe states are found at relatively weak
magnetic fields, we can anticipate that the valence LL
which is partially occupied will not be widely separated from 
remote LL's.  We include remote LL degrees of 
freedom in our calculation by accounting for the screening they
produce when polarized by valence LL electrons.  
The RPA 
(one-loop) calculation, leads to the following expression for the 
modified dielectric function\cite{notexact}:
\begin{equation}
\frac{\epsilon_(\vec q)}{\epsilon_0} =  
1 - \sum_{i',i,\sigma}^{'} 
\frac{ n_F(\varepsilon_{i',\sigma}) - n_F(\varepsilon_{i,\sigma})}
{2 \pi \ell^2 (\varepsilon_{i',\sigma} - \varepsilon_{i,\sigma}) } 
 \, V_{\sigma}^{i',i}(\vec q) \exp( -q^2 \ell^2/2) 
\label{epsilon}      
\end{equation}
where $\epsilon_0$ is the dielectric constant of the host semiconductor,
$n_F(x)$ is a Fermi factor,
the prime on the sum excludes the valence 
LL, and the effective 
inter-LL interactions
$V_{\sigma}^{i',i}(\vec q)$ 
differ from $V(\vec{q})$ only through the replacement
of $M_{\sigma}^{i,i}(\vec q)$ by $M_{\sigma}^{i',i}(\vec q)$.  
The wavefunctions and single-particle 
eigenvalues, $\varepsilon_{i,\sigma}$, 
used to define the effective interactions were obtained 
from local-spin-density self-consistent-field calculations which include  
the solution of the two-dimensional single-particle
Schr\"{o}dinger equation that arises\cite{tomasludwig} at tilted 
magnetic fields.
The effective interactions are anisotropic because $B_{\parallel}$ 
mixes the cyclotron and electric subband levels.

One-particle density matrices in a single LL, and 
hence also Hartree-Fock (HF) energies\cite{hfrefs}, are 
uniquely specified\cite{densitymatrix} by the particle density
function.  The energy per electron of the UCDW state 
at fractional filling $\nu^*$ of the valence LL
is given by \cite{hfrefs}
\begin{equation}
\label{ucdwen}
E=\frac{1}{2\nu^*}\sum_{n=-\infty}^{\infty} \Delta_n^2
U\left(\frac{2\pi n}{a}\hat{e}\right)\; ,
\Delta_n=\nu^*\frac{\sin(n\nu^*\pi)}{n\nu^*\pi}\; ,
\end{equation}
where $a$ is the period of the UCDW state and $\hat e$ is the 
direction of charge variation.
The UCDW state consists of stripes of width $a \nu^*$ with 
occupied guiding center
states separated by stripes of width $a (1 - \nu^*) $ 
with empty guiding center states; 
$\Delta_n$ above is the Fourier transform of the the guiding
center occupation function at wave vector $ n 2 \pi /a$.
In HF theory, the UCDW state energy depends only on
$a$ and $\hat e$ and the optimal UCDW is obtained by
minimizing Eq.(\ref{ucdwen}) with respect these parameters.
In Eq.(\ref{ucdwen}), $U(\vec{q})$ 
can be separated into  direct, $H(\vec{q})$, 
and exchange,
$X(\vec{q})$, contributions with 
\begin{eqnarray}
H(\vec{q})&=&\frac{1}{2\pi\ell^2}\, e^{-q^2\ell^2/2}\,
V(\vec{q}) \nonumber \\
X(\vec{q})&=&-\int\frac{d^2p}{(2\pi)^2}\, 
e^{-p^2\ell^2/2}\,
e^{i(p_xq_y-p_yq_x)\ell^2}\, V(\vec{p})\; .
\label{hartree}
\end{eqnarray}

The physics responsible for the occurrence of UCDW states is 
simple and robust.
For an infinitely narrow electron layer the effective 2D Coulomb 
interaction, $V(\vec{q})$, reduces to $(L_N(q^2/2))^2\, 
2\pi e^2\ell/\epsilon q$ where $L_N(x)$ is a 
Laguerre polynomial. 
Starting from $N=1$, zeros of $L_N(x)$ occur at 
smaller $x$ with increasing $N$, producing a zero in the repulsive Hartree
interaction at smaller wave vector where the attractive
exchange interaction is stronger.  
In Table I we compare the $\nu^*=1/2$ HF energies of 
triangular Wigner crystal states and UCDW states with maximum $a$ 
satisfying $H(2 \pi/a ) =0$.  The triangular 
Wigner crystal state energy is intended to approximate the energy of 
possible isotropic fluid states.  We see that for $N \ge 2$, 
the energetic preference for the UCDW states is large, substantially
larger for example than the preference for Laughlin's\cite{rbl} fluid 
states over Wigner crystal states at $\nu = 1/3$.  These calculations
suggest that for $N=1$ the competition between isotropic fluid states
and UCDW states is delicate.  Also noted in Table I is the fact
that in the HF approximation, the UCDW state is unstable
to charge modulation along the stripes\cite{fradkin}, leading to anisotropic
Wigner crystal states with slightly lower energy.  
This instability is, however, misrepresented by the HF approximation
and the system is expected\cite{mpaf,fradkin} to be effectively
a UCDW at any accessible temperature for $0.4 < \nu^* < 0.6$.
We appeal to the relatively small difference 
between UCDW and anisotropic Wigner crystal state HF energies
in using the simple UCDW state to estimate the anisotropy energy.

We now turn to our evaluation for the anisotropy energy at 
$\nu$=5/2, 9/2, and 13/2 in the sample of
Lilly {\em et al.}\cite{lilly,lilly2}.  The self-consistent-field
separation between lowest spin-up electrical subbands is 9.8~meV
so that the valence LL's at perpendicular field for these filling 
factors are 
the spin-up N=1, 2, and 3 LL's of the first electrical 
subband, respectively.  The in-plane magnetic field has only a weak
effect on the LL spacing even at field-tilt angles as high as $\theta
=60^o$.
We represent the effective interaction anisotropy by performing a 
Fourier expansion  in the angle $\phi$ between $\hat e$ and the 
in-plane field:
\begin{eqnarray}
\label{hartreefou}
H(q,\phi)&=&\sum_n H_{2n}(q)\cos(2n\phi)
\nonumber \\
X(q,\phi)&=&\sum_n X_{2n}(q)\cos(2n\phi)\; ,
\end{eqnarray}
where 
\begin{equation}
X_{2n}(q)=-\int_0^{\infty}dp\,pH_{2n}(p)J_{2n}(pq)
\end{equation}
and $J_{m}(x)$ is the Bessel function. 
Even at large $B_{\parallel}$
the anisotropy of the effective interaction is relatively weak and 
is accurately proportional to $\cos(2\phi)$.
This property of $H(\vec{q})$ is shared by $U(\vec{q})$ and 
greatly simplifies the UCDW energy~(\ref{ucdwen}) minimization procedure.
For a given $a$ the extrema of $E$ lies either at  
$\phi=0$ or at $\phi=\pi/2$. We define the anisotropy energy per
electron $E_A$
as the  minimum of $E(\phi=\pi/2)$ minus the minimum of $E(\phi=0)$. 

Details of the anisotropy energy
calculation are summarized in Table II.  We first discuss the results
obtained when RPA screening is neglected.  Most qualitative features
are already captured in a simple theory which retains only the 
$n=1$ leading harmonic in the UCDW energy expression and finds the 
optimal UCDW period $a=a^*_0$ by minimizing $H_0(2\pi/a)+X_0(2\pi/a)$.  
The Hartree
anisotropy energy $E^H_{A,0}=-2H_2(2\pi/a^*_0)$ is consistently
negative (stripes along in-plane field) but is 
countered by the exchange energy $E^X_{A,0}=-2X_2(2\pi/a^*_0)$.
For $\nu=9/2$ and 13/2, 
where the UCDW state is most robust, the Hartree
term dominates when screening is neglected but exchange dominates 
when  screening is  accounted for.
Our finding that the stripes prefer to be aligned
perpendicular to the $B_{\parallel}$ direction is consistent with
the experimental finding\cite{lilly2,panetal} that this is the easy
transport direction.  Including all harmonics in the UCDW energy
expression and reoptimizing the lattice constant $a=a^*$, 
substantially
reduces  numerical value of the anisotropy energy
but does not change its sign.  $E_A$ 
is largest in magnitude for $\nu=5/2$.
Even these relatively modest anisotropy energies are sufficient
to tip the delicate balance between isotropic and anisotropic states
for $N=1$, explaining the transition to anisotropic states 
seen in experiment.  We can use the calculated values for $E_A$ to
estimate the temperature below which anisotropy will be observed in
the transport properties of these systems.
Current experimental samples
apparently have a native anisotropy of unknown 
origin which can be overcome by the application of an
in-plane field, reorienting the stripes and changing the easy transport
direction.   Since $\theta<20^0$ can reorient the 
stripes for $N=2$ and $N=3$, we estimate from Table II that the native 
anisotropy energy is less than $10^{-4} (e^2/\epsilon_0 \ell)
\sim k_B 10 {\rm mK}$ per electron.
We can also use $E_A$ to
estimate the temperature below which anisotropy will be observed in
the transport properties of these systems.  Based on an
experimental onset temperature $T^* \sim 100 {\rm mK}$ with native anisotropy
we estimate that $k_B T^* \sim 10 E_A$.  According to our calculations
the largest anisotropies occur for $N=1$ for which we predict 
an onset temperature exceeding 1K at large $\theta$.
We note that our theory gives similar results for the field-tilt anisotropy 
energy at $\nu=11/2$ and $\nu=9/2$ and therefore as unable to explain 
the differences observed in the anisotropic transport 
measurements\cite{lilly2} in 
majority and minority valence LL's. 

\begin{figure}[b]
\epsfxsize=3.5in
\centerline{\epsffile{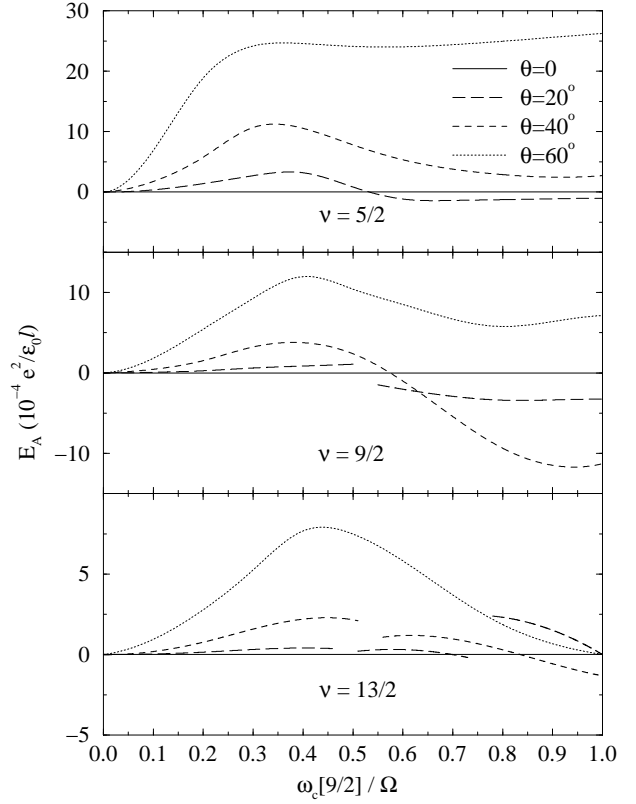}}

\vspace*{.0cm}

\caption{Field-tilt anisotropy energy as a function of parabolic confining 
potential strength.
Data for the valence LL close to degeneracy with another LL are not 
plotted as the theory fails to describe this circumstance.}
\label{fig1}
\end{figure}

\begin{figure}[b]
\epsfxsize=3.5in
\centerline{\epsffile{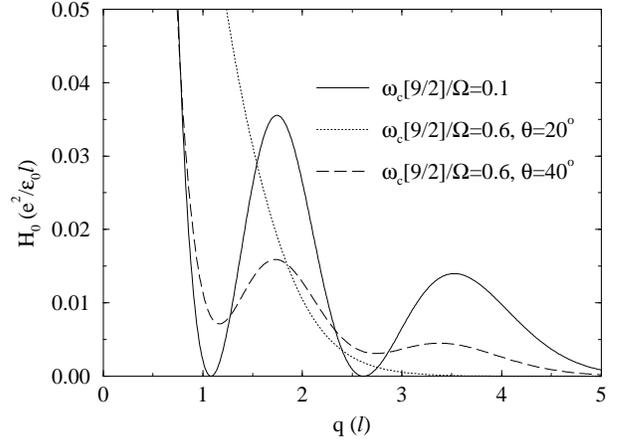}}

\vspace*{.0cm}

\caption{Wavevector dependent Hartree energies for parabolic quantum well
model and $\nu=9/2$.}
\label{fig2}
\end{figure}

Finally, we discuss  UCDW energy calculations 
for parabolic quantum wells with different
electric subband splittings $\hbar\Omega$.
The results are summarized in Fig.~\ref{fig1}; 
both screening and higher harmonics in the 
UCDW energy were accounted for in these calculations. 
The perpendicular magnetic field was chosen to correspond to 
the 2D electron density in the experiments of  
Lilly {\em et al.}\cite{lilly,lilly2}, {\em i.e.}
the cyclotron frequency at $\nu=9/2$ is $\hbar \omega_c[{9/2}]=4.24$~meV.
Two regimes can be distinguished in Fig.~\ref{fig1}.  For narrow
quantum wells ($\omega_c[{9/2}]/
\Omega < 0.5$), only the lowest electrical subband is occupied at 
$\theta=0$, the stripes orient perpendicular to the in-plane field,
and the magnitude of $E_A$ increases with $\theta$ and decreases with $N$.
The samples of Lilly {\em et al.}\cite{lilly,lilly2} and 
Pan {\em et al.}\cite{panetal}  
fall into this regime.  In wider quantum wells the perpendicular
field valence LL can belong to a higher electrical subband,
and more complex behavior occurs. 
The solid curve in Fig.~\ref{fig2} shows the interaction energy 
$H_0$ for $\nu=9/2$ and a narrow parabolic quantum
well. It has a structure characteristic of the $N=2$ LL effective
interaction. 
($\theta$ is not indicated here 
as the field-tilt has a negligible effect on
$H_0(q)$ for $\omega_c[{9/2}]/\Omega=0.1$.) The dotted and dashed
curves correspond
to the case where the perpendicular field valence LL 
is the lowest ($N=0$) LL of the second
electrical subbands. For $\theta=20^o$, $H_0(q)$ decreases monotonically
with $q$, as in the perpendicular field;
as explained above the UCDW is not the likely ground state
for the system in this circumstance.  However, at $\theta=40^o$,
$H_0(q)$  is more akin the perpendicular field
$N=2$ LL effective interaction which favors  the UCDW state.
This mechanism of stabilizing UCDW ground state by in-plane 
magnetic field is different from the one discussed above for
Lilly's {\em et al.}\cite{lilly,lilly2} sample and is germane 
to wider quantum wells with higher electrical subbands occupied.
Our calculations indicate that both perpendicular and
parallel orientations of the stripes with respect to 
the in-plane field can be realized for these tilt-induced UCDW states.
The competition between isotropic and anisotropic states, and 
the anisotropy energy of UCDW states, will both have a 
complicated dependence on filling factor and tilt-angle in this
regime. 

The authors acknowledge stimulating interactions
with J. P. Eisenstein and thank M. Fogler and R. Moessner 
for helpful private communications. 
This work was supported by NSF  
grant DMR-9714055 
and by the Grant Agency of the Czech Republic
under grant 202/98/0085.

\begin{table}
\begin{center}
\begin{tabular}{ccccc}
$N$ & $e^{TWC}$
& $a/\ell$ & $e^{UCDW}$ & $e^{AWC}$
\\ \hline
0 & -0.4435 & 3.299 & & \\
1 & -0.3443 & 4.443 & -0.3456 & -0.3509 \\  
2 & -0.2897 & 5.805 & -0.3063 & -0.3091 \\
3 & -0.2667 & 6.890 & -0.3041 & -0.3061 
\end{tabular}
\end{center}
\caption{\protect
HF state energies per electron at $\nu^*=1/2$ for triangular Wigner crystal,
UCDW, and anisotropic Wigner crystal states. The energies are in units
of $e^2/\epsilon_0 \ell$.  These results are for zero
thickness 2D electron layers and no screening.}
\label{energies}
\end{table} 

\vspace*{-0.0cm}

\begin{table}
\begin{center}
\begin{tabular}{rrrr|rrrr}
 &
 \multicolumn{3}{c|}{NO SCREENING} & 
 \multicolumn{4}{c}{SCREENING} \\
 $\theta$  &  $E_{A,0}^H$ & $E_{A,0}^X$ & $E_{A}$ & 
 $E_{A,0}^H$ & $E_{A,0}^X$ & $E_{A}$ & $a^*/\ell$ \\ \hline
\multicolumn{8}{c}{$\nu=5/2$} \\
$20^o$ & -32.79 & 36.16 & 1.16 & -17.65 & 28.94 & 2.80 & 5.15 \\
$40^o$ & -45.70 & 78.73 & 8.85 & -21.26 & 70.63 & 12.38 & 5.24 \\
$60^o$ & -127.32 & 174.73 & 10.73 & -75.64 & 164.19 & 21.25 & 5.15 \\
\hline
\multicolumn{8}{c}{$\nu=9/2$} \\
$20^o$ & -13.52 & 6.17 & -1.40 & -5.58 & 7.59 & 0.27 & 6.41 \\
$40^o$ & -43.84 & 18.17 & -4.44 & -10.83 & 19.51 & 2.23 & 6.41 \\
$60^o$ & -101.57 & 39.78 & -9.59 & -15.00 & 47.00 & 8.07 & 6.68 \\
\hline
\multicolumn{8}{c}{$\nu=13/2$} \\
$20^o$ & -3.76 & 0.06 & -0.77 & -0.81 & 0.91 & 0.04 & 7.66 \\
$40^o$ & -18.49 & 2.75 & -3.48 & -1.93 & 6.20 & 0.87 & 7.66 \\
$60^o$ & -70.55 & 6.33 & -12.49 & -2.54 & 16.03 & 2.68 & 7.85 
 \end{tabular}
\end{center}
\caption{\protect
Field-tilt anisotropy energy components.  Energies are per electron
and in units of 
$10^{-4}$  $e^2/\epsilon_0\ell \sim k_B 10 {\rm mK} $.}
\label{tab}
\end{table}


\begin{references}
\bibitem{caveats} This statement applies when mixing between
different LL's is neglected.
\bibitem{notexact} Because 2D and subband response is coupled in
a tilted field, this is the exact RPA expression only for zero
tilt angle.  The field-tilt dependence of remote LL
screening is not important in our calculations, however.
\end{references}
\end{document}